\documentclass[12pt]{iopart}

\usepackage{iopams}  
\usepackage{graphicx}
\usepackage{amsbsy}

\usepackage[usenames]{color}
\usepackage[normalem]{ulem}

\begin{document}

\title{Fluctuating hydrodynamics and correlation lengths in a driven granular fluid}

\author{G. Gradenigo, A. Sarracino, D. Villamaina and A. Puglisi}
\address{CNR-ISC and Dipartimento di Fisica, Universit\`a Sapienza - p.le A. Moro 2, 00185, Roma, Italy}

\ead{ggradenigo@gmail.com,alessandro.sarracino@roma1.infn.it,
\\ dario.villamaina@roma1.infn.it,andrea.puglisi@roma1.infn.it}

\begin{abstract}
Static and dynamical structure factors for shear and longitudinal modes
of the velocity and density fields are computed for a granular system
fluidized by a stochastic bath with friction. Analytical expressions
are obtained through fluctuating hydrodynamics and are successfully
compared with numerical simulations up to a volume fraction $\sim
50\%$. Hydrodynamic noise is the sum of external noise due to the bath
and internal one due to collisions. Only the latter is assumed to
satisfy the fluctuation-dissipation relation with the average granular
temperature. 

Static velocity structure factors $S_\perp(k)$ and $S_\parallel(k)$
display a general non-constant behavior with two plateaux at large and
small $k$, representing the granular temperature $T_g$ and the bath
temperature $T_b>T_g$ respectively. From this behavior, two different
velocity correlation lengths are measured, both increasing as the
packing fraction is raised.  This growth of spatial order is in
agreement with the behaviour of dynamical structure factors, the decay
of which becomes slower and slower at increasing density.  
\end{abstract}

\pacs{45.70.-n, 51.20.+d, 05.40.-a, 47.57.Gc}

\maketitle

\section{Introduction}
\label{intro}

Granular media display a wide catalog of non-equilibrium
phenomena~\cite{JNB96b}. These materials are constituted by a number
$N$ of elementary constituents, grains of typical diameter between
$0.1$ and $10$ mm. The number $N \gg 1$ is usually large enough to
allow, or require, a statistical treatment. Unfortunately,
interactions are non-conservative, resulting in the failure of
equilibrium statistical mechanics. Kinetic theories, from Boltzmann
equation to hydrodynamics~\cite{PL01}, together with numerical
simulations~\cite{PS05}, are the best tools to describe those systems
and to compare with real experiments, with the {\em caveat} of a
proper adaptation to the peculiarity of granular interactions.

One of the debated points of granular kinetic theories is the way
noise should be added to hydrodynamics in order to describe mesoscopic
fluctuations~\cite{LandauFisStat,FU70}. This is a general problem in
non-equilibrium systems~\cite{OZS06} (e.g. sheared
fluids~\cite{SOZ10}), but here is even more pressing, given the rather
small number of particles in a granular system: one has typically $N
\sim 10^3\div 10^4$, even in experiments, so that fluctuations can
hardly be neglected. Moreover, in granular systems the dynamics is
non-conservative and therefore Fluctuation-Dissipation relations do
not hold in general~\cite{PBV07,BGM08,VPV08,SVGP10}, with exceptions
in driven dilute cases~\cite{PBL02,G04,SVCP10}. In non-dilute systems,
it is therefore difficult even to define a temperature, making tricky
the modelization of fluctuations~\cite{DMGBLN03,BBDLMP05}.

A comprehensive study of the fluctuating hydrodynamics of a driven
granular fluid is presented here. Static and dynamical structure factors
are computed analytically in the framework of linearized hydrodynamics,
and compared with extensive numerical simulations. A very good
agreement is found between analytical and numerical results in a wide
range of parameters, implying that, for this kind of model,
fluctuating hydrodynamics is able to describe large scale fluctuations
in a satisfactory manner.

The peculiarity of the model we have studied, when compared to others
present in the literature~\cite{WM96,NETP99}, is the prescription for
the stochastic bath used to keep the system at stationarity.  In
particular our thermostat is able to equilibrate the system also when
collisions are elastic~\cite{PLMPV98}.  This happens because, in
addition to a random driving, our thermostat acts on the particles
also through a finite drag, in such a way that the temperature of the
thermostat $T_b$, different form the kinetic temperature of the fluid
$T_g<T_b$, is always well defined.

A remarkable feature of our model, related to the kind of thermostat
we use, is the finite extent of velocity correlations in space.
Indeed, the characteristic shape we find for transverse and
longitudinal velocity structure factors allows us to define two
non-equilibrium correlation lengths, $\xi$ and $\xi_l$, which are
known functions of the kinematic and longitudinal viscosities,
respectively. This means that, instead of sampling trajectories of the
system, out of equilibrium an average of static observables is enough
to measure transport coefficients.  The flattening of the velocity
structure factors at equilibrium clearly results in our formulas from
the vanishing of the velocity correlations amplitude, which is
proportional to $T_b-T_g$.  In that case, access to transport
coefficients is only possible through the study of the
\emph{dynamical} structure factors.

Finally, we study the behaviour of such coherence lengths at different
packing fractions. We observe a significantly growth of the relative
extent of correlations $\xi/\lambda_0$ and $\xi_l/\lambda_0$ with the
packing fraction, where $\lambda_0$ denotes the mean free path.  
This is in agreement with the slowing down of the dynamics 
also observed in dense granular fluids~\cite{VAZ11}.

The paper is organized as follows. In Section~\ref{Microscopicmodel}
we discuss the model, the corresponding hydrodynamic equations and the
specific forms of noise used. The $k$ dependence of the eigenvalues of
the linearized hydrodynamic matrix is also studied in order to check
the stability and bounds of the linear approximation. In
Section~\ref{Out-of-equilibrium} we present a complete study of the
static and dynamical structure factors for the velocity and density
modes, comparing analytical predictions and numerical results. In
section~\ref{Summary} a discussion about the main findings is
presented, with conclusions and perspectives for future
work. In~\ref{Enskog} and~\ref{app} are reported, respectively,
formulas for the transport coefficients and some details on the noise
terms.

\section{Microscopic model and fluctuating hydrodynamics} 
\label{Microscopicmodel}

We consider a system of $N$ inelastic hard spheres in $d$ dimensions
with mass $m$ and diameter $\sigma$.  Particles are contained in a
volume $V=L^d$, with $L$ the linear size of the system.
We denote by $n=N/V$ the number density and by $\phi$ the occupied
volume fraction, (in two dimensions $\phi=N\pi(\sigma/2)^2/V$).  The
particles undergo binary instantaneous inelastic collisions when
coming at contact, with the following rule
\begin{equation} 
{\bf v}_i={\bf v}'_i-\frac{(1+\alpha)}{2}
\left[\left({\bf v}'_i-{\bf v}'_j\right)
\cdot\hat{\boldsymbol{\sigma}}\right]\hat{\boldsymbol{\sigma}} 
\label{m.1}
\end{equation}
where ${\bf v}_i$ (${\bf v}_j$) and ${\bf v}'_i$ (${\bf v}'_j$) are
the post and pre-collisional velocities of particle $i$ (particle
$j$), respectively; $\alpha\in[0,1]$ is the restitution coefficient
(in the elastic case $\alpha=1$), and $\hat{\boldsymbol{\sigma}}$ is
the unit vector joining the centers of the colliding particles.

In order to maintain a stationary fluidized state, an external energy
source is coupled to every particle in the form of a thermal bath~\cite{PLMPV98}. In
particular, the motion of a particle $i$ with velocity ${\bf v}_i$ is
described by the following stochastic equation
\begin{equation}
m\dot{{\bf v}_i}(t)=-\gamma_b {\bf v}_i(t) + \boldsymbol{\xi}_{b,i}(t)+{\bf F}_i.
\label{m.2}
\end{equation}
Here $\gamma_b$ is a drag coefficient (which defines the
characteristic interaction time with the external bath,
$\tau_b^{-1}=\gamma_b/m$), $\boldsymbol{\xi}_{b,i}(t)$ is a white
noise with $\langle\boldsymbol{\xi}_{b,i}(t)\rangle=0$ and
$\langle\xi_{b,i\alpha}(t)\xi_{b,j\beta}(t')\rangle=
2T_b\gamma_b\delta_{ij}\delta_{\alpha\beta}\delta(t-t')$ (Greek
indexes denote Cartesian coordinates), while ${\bf F}_i$ represents
the action of particle-particle inelastic collisions.  The effect of
the external energy source balances the energy lost in the collisions
so that an \emph{out-of-equilibrium} stationary state is
attained~\cite{PLMPV98}. In particular, let us stress
  that such energy injection mechanism acts homogeneously across the
  whole system, differently from other mechanisms where the energy is
  directly supplied only to a part of the system, as for instance for
  fluids under shear, or for systems in conctact with vibrating
  walls.

The stationary state is characterized by two time scales and two
energy scales: the time scales are $\tau_b$ and the mean free time
between collisions $\tau_c$; the energy scales are the temperature of
the thermostat $T_b$ and the granular temperature
$T_g=\frac{\sum_{i=1}^{N} m\langle v_i^2\rangle}{dN} \le T_b $ (equal
sign holds only if $\alpha=1$ or if $\tau_b \ll \tau_c$).  In
particular, in the dilute limit, the granular temperature $T_g$
satisfies the following equation in the non-equilibrium stationary
state \cite{SVCP10}
\begin{equation}
T_g=T_b-A_d\frac{\chi(\phi)\phi(1-\alpha^2)}{2\gamma_b}T_g^{3/2},
\label{tg}
\end{equation}
where $A_d=\sqrt{m/\pi}2^{d-1}d/\sigma$ and $\chi(\phi)$ is the pair
correlation function at contact.  The model has a well-defined elastic
limit $\alpha\to 1$, where the fluid equilibrates to the bath
temperature, $T_g=T_b$.  The viscous drag term $-\gamma_b{\bf v}_i$ in
Eq.~(\ref{m.2}) models the interaction between each particle and the
thermostat.  It is important to observe that $\gamma_b$ is not related
to the transport coefficients of the granular fluid and is fixed as a
model parameter. As mentioned above, $\gamma_b$ introduces a time
scale $\tau_b$ in the system that rules the tendency of particle to
relax toward equilibrium at temperature $T_b$. The characteristic time
of collisions, $\tau_c$, in all our simulations will be kept much
smaller than $\tau_b$: for this reason $\tau_c$ is considered the
\emph{microscopic} time-scale of our system since it dictates the
smallest scale of relaxation toward the non-equilibrium stationary
state. In particular, in the coarse-grained hydrodynamic description
to be introduced below, we will take care of comparing the
characteristic decay time of different hydrodynamics modes with
$\tau_c$, to verify the presence of a sufficient separation of scales.

In the following we will present a thorough numerical analysis of
model~(\ref{m.2}), using an event-driven molecular dynamics
algorithm~\cite{l05}.  In particular, we will consider periodic
boundary conditions in $d=2$ dimensions.  The fixed parameters of the
simulations are $m=1$, $\sigma=0.01$, $T_b=1$ and $\gamma_b=1$. The
packing fraction is varied by changing the seize of the box, and we
consider systems with $\phi\in[0.1,0.5]$.  The simulation data on
static structure factors are obtained for a system of $N=10000$
particles, averaged over about 100 realizations, whereas the results
on dynamical correlators are obtained with samples of $N=1000$
particles, averaged over about 4000 realizations.

\subsection{Linearized hydrodynamics}
\label{Fluctuations}

Because we are interested in the behavior of large-scale spatial
correlations in our system, we introduce here the coarse-grained
hydrodynamic fields $n({\bf r},t),{\bf u}({\bf r},t)$ and $T({\bf
  r},t)$ as follows:
\begin{eqnarray}
n({\bf r},t)&=&\sum_i\delta({\bf r}-{\bf r}_i(t)), \nonumber \\
{\bf u}({\bf r},t)&=&\frac{1}{n}\sum_i {\bf v}_i(t)\delta({\bf r}-{\bf r}_i(t)), \label{he.0} \\
T({\bf r},t)&=&\frac{2m}{dn}\sum_i \frac{v^2_i(t)}{2}\delta({\bf r}-{\bf r}_i(t)). \nonumber 
\end{eqnarray}
The hydrodynamic equations for the fields~(\ref{he.0}) can be derived
for the model~(\ref{m.2}) following a standard
recipe~\cite{F75,GZB97,BDKS98}:
\begin{eqnarray}
\partial_{t}n({\bf r},t)&=& - \boldsymbol{\nabla}\cdot(n({\bf r},t){\bf u}({\bf r},t))
\nonumber \\ 
\partial_{t}{\bf u}({\bf r},t) + {\bf u}\cdot\boldsymbol{\nabla}{\bf
  u} &=& - \frac{1}{\rho}\boldsymbol{\nabla}\cdot\boldsymbol{\Pi} - \frac{\gamma_b}{m}{\bf
  u}({\bf r},t) \label{he.1}\\ 
\partial_t T({\bf r},t) + {\bf u}\cdot\boldsymbol{\nabla}T
&=& -\frac{2}{nd}(\boldsymbol{\nabla}\cdot {\bf J}+ \boldsymbol{\Pi}:\boldsymbol{\nabla}{\bf u}) 
-\Gamma + 2\frac{\gamma_b}{m}(T_b - T({\bf r},t)) . \nonumber
\end{eqnarray}
In the above equations ${\bf J}$ and $\boldsymbol{\Pi}$ are
respectively the heat flux and the pressure tensor, see details
in~\ref{app}, and $\gamma_0=(1-\alpha^2)/2d$. In the velocity equation
the viscous drag term $-\gamma_b{\bf u}/m$ has been inserted, while in
the temperature equation three terms have been added: the sink term
$-\Gamma=-2\gamma_0\omega_cT({\bf r},t)$~\cite{NEBO97}, where
$\omega_c\sim\sqrt{T(\bf{r},t)}$ is the collision frequency, takes into account
the energy dissipated by inelastic collisions, while the terms
$2\gamma_b(T_b-T)/m$ represent the energy exchanged with the
thermostat.

Eqs.~(\ref{he.1}) give a fair description of the mesoscopic degrees of
freedom of a granular fluid as long as a proper separation of space
and time scales is verified between those degrees of freedom and all
the microscopic ones which are projected out. This condition is, of
course, not always satisfied~\cite{G99,K99,PCV05}, but is not
prevented in principle and is, indeed, realized in many experiments or
simulations~\cite{HMD02,BP05,PAFM08,VAZ11,GSVP11}.

Eqs.~(\ref{he.1}) can be linearized around the stationary
\emph{homogeneous} state, where the hydrodynamic fields take the
values $n=\overline{n}$, $T=\overline{T}$ and ${\bf u}=0$. A system of
linear differential equations for the fluctuations $\delta{\bf a}({\bf
  k},t)=\{\delta n({\bf k},t),\delta T({\bf k},t),u_\parallel({\bf
  k},t),u_\perp({\bf k},t)\}$, with $\delta a=a-\overline{a}$, can be
considered, with the Fourier transform defined as
\begin{equation}
\delta {\bf a}({\bf k},t)=\int d{\bf r}~\delta {\bf a}({\bf r},t)e^{- i{\bf k}\cdot{\bf r}},
\label{ft.1}
\end{equation} 
and with $u_\perp({\bf k},t)$ and $u_\parallel({\bf k},t)$
respectively the shear and longitudinal modes, namely
\begin{eqnarray}
u_\parallel({\bf k}) &=& \hat{k}\cdot {\bf u}({\bf k}) \nonumber \\
u_\perp({\bf k}) &=& \hat{k}_\perp \cdot {\bf u}({\bf k}), 
\end{eqnarray}
$\hat{k}_\perp$ being a unitary vector such that $\hat{k}_\perp \cdot
\hat{k}=0$.  The system in Eq.~(\ref{he.1}) in Fourier space becomes
\begin{equation}  
\delta\dot{\bf a}({\bf k},t)={\bf M}(k)\delta{\bf a}({\bf k},t),
\end{equation}
with the dynamical matrix

\begin{equation}
\hspace{-2.5cm}\mathbf{M}(k) = -
\left( \begin{array}{cccc}
0 & 0 & \imath kn & 0 \\
\gamma_0 \omega_c g(n) T_g/n & 3\gamma_0 \omega_c +D_T k^2 + 2\gamma_b/m &  \imath 2kp/dn & 0 \\
\imath k v_T^2/n & \imath kp/\rho T_g &  \nu_l k^2 +\gamma_b/m & 0 \\
0 & 0 & 0 & \nu k^2+\gamma_b/m 
\end{array} \right),
\label{fh.1}
\end{equation}
where $\rho=n m$, $D_T=2 \kappa/n d$ is the thermal diffusion
coefficient ($\kappa$ is the heat conductivity), while $\nu$ and
$\nu_l$ are the kinematic and longitudinal viscosity
respectively. Formulas for all parameters and transport coefficients
are given in~\ref{Enskog}. There we refer to the Enskog
  theory for dense elastic hard spheres (EHS)~\cite{CC70}, which
  provides a good approximation, as observed in~\cite{NETP99,GLB06}.
The following sections are devoted to show how the viscosities $\nu$
and $\nu_l$ can be obtained as fit parameters of static and dynamical
correlations. Such results will be compared with the dense EHS
predictions, finding good agreement.

\subsection{Spectrum of the hydrodynamic matrix and separation of time-scales}
\label{Spectrum}

\begin{figure}[tbp!]
\begin{center}
\includegraphics[width=0.6\columnwidth,clip=true]{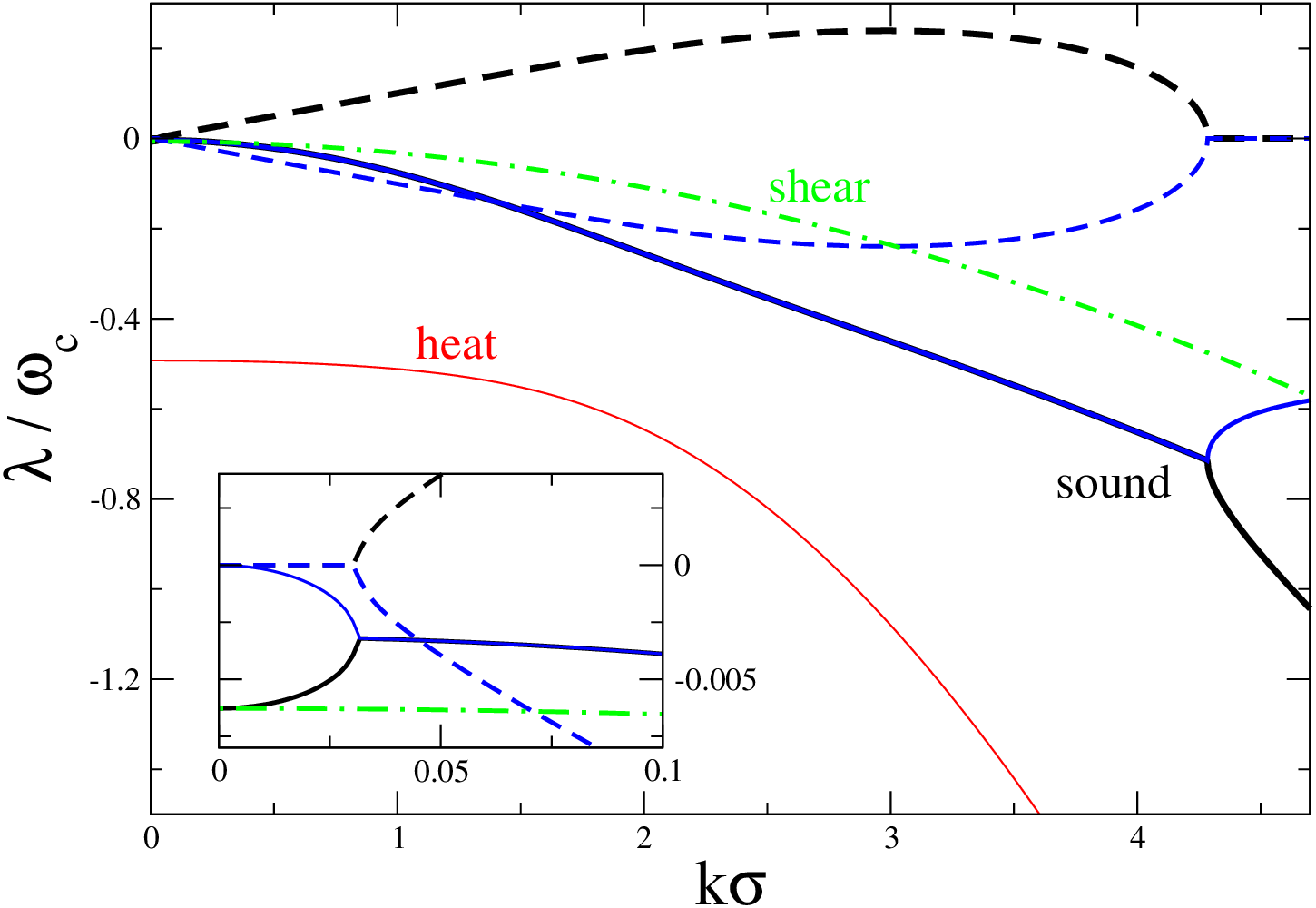}
\caption{Eigenvalues spectrum $\lambda(k)$ of the dynamical matrix~(\ref{fh.1})
  calculated for $\phi=0.5$ and $\alpha=0.6$, namely in a strongly
  inelastic regime.  The eigenvalues are normalized on the Enskog
  collision frequency $\omega_c$ so that on $y$ axis the ratio
  $\tau_c/\tau(k)=\lambda(k)/\omega_c$ between the microscopic
  time-scale of the model and the decay time of each mode can be
  directly read. Inset: zoom of the spectrum at small $k$.}

\label{eigenvalues}
\end{center}
\end{figure}      
We analyze the eigenvalues of ${\bf M}(k)$ in order to study the
linear stability of the model and to characterize the range of
validity of time-scales separation required by hydrodynamics. From the
expression in Eq.~(\ref{fh.1}) we learn that the shear modes are
decoupled from all the others, and the typical time-scale of their
decay simply reads as $\tau_\perp = ( \gamma_b/m + \nu k^2 )^{-1} $.
To obtain the typical time-scales for the fluctuations of the other
hydrodynamic fields we solve, numerically, the equation
$\textrm{Det}({\bf M}(k)-\lambda(k){\bf I})=0$, where ${\bf I}$ is the
identity matrix.  The eigenvalues spectrum thus found is shown in
Fig.~\ref{eigenvalues}, for parameters of the model $\alpha=0.6$ and
$\phi=0.5$. All parameters are calculated according to the formulas
reported in~\ref{Enskog}.

The study of hydrodynamic eigenvalues shows that, even in the case of
a quite inelastic and dense regime, a range of scales where mesoscopic
relaxation times are larger than the microscopic times exists and a
hydrodynamic description can be attempted.  In particular, comparing
the eigenvalues $\lambda_i(k)$ with the Enskog frequency $\omega_c$
(which will be verified, below, to be a good estimate of the real
frequency in simulations), see Fig.~\ref{eigenvalues}, we find, for
each eigenvalue, the interval of values of $k$ where
$\lambda_i/\omega_c < 1$ is fulfilled. The subscript $i$ indicates
eigenvalues related to heat ($i=H$), sound ($i=\pm$) and shear ($i=\perp$) modes.
The range is narrower for the case of the mode
dominated by temperature fluctuations (denominated ``heat mode'' in
the Figure, red thin curve), while it is larger for shear modes (green
dot-dashed lines) and modes dominated by longitudinal velocities and
density fluctuations (here referred to as ``sound modes'', black and
blue thick curves, continuous for real part and dashed for imaginary
part).

A general observation is that eigenvalues never have a positive real
part, i.e.  {\em no instabilities} are found, thanks to the presence
of the external bath. Moreover, for the whole range of $k$ studied,
the two ``sound modes'' - as usual - are complex conjugate, i.e. they
propagate with a $k$-dependent velocity. A negative real part is
always present, determining overall damping. At small values of $k$
one can identify a sound velocity $c$ by observing a linear relation
$\textrm{Im}[\lambda_{\pm}] \sim c k$; instead the dispersion relation
becomes strongly nonlinear at large $k$.  Surprisingly, the study of
sound eigenvalues shows that they have bifurcations at very small
$k<k_1$ and very large $k>k_2$ wave-numbers, becoming in both limits
pure real numbers, i.e. losing their propagating behavior. Those
bifurcations are due to the external damping, ruled by $\gamma_b$:
indeed in the limit $\gamma_b \to 0$ (keeping finite $\gamma_bT_b$),
they disappear and the spectrum studied in~\cite{NETP99,VAZ11} is
retrieved. Moreover, we find that the wavevector $k_1$ where
  the first bifurcation occurs, moves towards smaller values of $k$ when the
  dissipation is increased, namely when $\alpha$ is decreased (at
  fixed $\phi$) or when $\phi$ is increased (at fixed $\alpha$).
Looking closely to the bifurcations, we see that one of the
eigenvalues for ``sound modes'' approaches $0$ for $k \rightarrow 0 $
(representing total number conservation) and the other tends to
$-\gamma_b/m$ (as the shear one), see inset of
Fig.~\ref{eigenvalues}. The bifurcation at large $k$ is perhaps
non-physical, as it always falls out of the hydrodynamic range.  For
the eigenvalue of the heat mode we find that $\lambda_H(k = 0)=-
(3\gamma_0\omega_c+2\gamma_b/m)$.  In the numerical setup used below,
we have $k_{min}=2\pi/L>k_1$, so that the purely exponential decay of
sound modes with large waves is never observed.

\subsection{Stochastic description with fluctuating hydrodynamics}
\label{SmallScale} 

In order to fully account for the spatial \emph{fluctuations} of the
hydrodynamic fields and for the decay in time of such fluctuations we
must add some noise terms to the linearized hydrodynamic equations:
the basic assumption under fluctuating hydrodynamics is the same as
for average (deterministic) hydrodynamics, i.e. a good separation of
scales between hydrodynamic fields and microscopic degrees of
freedom. 
In the linearized hydrodynamic equations the small
scale fluctuations have been projected out, but their feedback on large scale
fluctuations can be recovered by a proper addition of noise terms to dynamical
equations:
\begin{equation}
\delta\dot{\bf a}({\bf k},t)={\bf M}(k)\delta{\bf a}({\bf k},t) + {\bf f}({\bf k},t).
\label{fh.2}
\end{equation}
A derivation from first principles of the noise ${\bf f}({\bf k},t)$
is beyond our scope. A kinetic theory with fluctuations has been
recently proposed in~\cite{BMG09,BMG11}, for the homogeneous cooling
regime, which is very different from our case. A similar treatment has
been realized, only for the shear mode, in a driven case (random kicks
without damping, i.e. $\gamma_b=0$)~\cite{MGT09}, showing that noise
can be safely assumed to be white, at difference with the cooling
regime. In such a case, with an additional but reasonable assumption
on the two-particle velocity autocorrelation functions (namely, that
such functions have only components in the hydrodynamic subspace), it
is found that the fluctuation-dissipation relation for the internal
part of the noise is satisfied, as already assumed
in~\cite{NEBO97,NETP99}. Following those previous studies, we will
consider valid such an assumption. Notice that it does not imply that
fluctuation-dissipation relations will be satisfied by the whole
noise, which is composed by internal as well as {\em external}
contributions. In summary we write
\begin{equation}
{\bf f}({\bf k},t) =
\left( \begin{array}{c}
0  \\
\theta^{ex}({\bf k},t)+2ik/nd~\theta^{in}({\bf k},t) \\
\xi_l^{ex}({\bf k},t)+ik/\rho~\xi_l^{in}({\bf k},t) \\
\xi_\perp^{ex}({\bf k},t)+ik/\rho~\xi_\perp^{in}({\bf k},t)\}
\end{array} \right),
\label{fh.2bis}
\end{equation}
where the two sources of noise for the hydrodynamic fields
fluctuations are put in evidence: the first is the \emph{external}
contribution coming directly from the thermal bath, namely the
stochastic force $\boldsymbol{\xi}_b$ of Eq.~(\ref{m.2}); the second
is \emph{internal} and enters through the constitutive equations for
the heat flux and the pressure tensor.  A detailed discussion on
noises is presented in~\ref{app}.  The external and internal noises
are Gaussian with zero average.  The variances of external noises can
be obtained directly from Eqs.~(\ref{m.2}) and~(\ref{he.0})

\begin{eqnarray}
\langle\theta^{ex}({\bf k},t)\theta^{ex}({\bf k}',t')\rangle &=& \frac{4mT_g}{dn}\frac{2\gamma_bT_b}{m} 
\delta(t-t')\delta({\bf k}+{\bf k}') \nonumber  \\
\langle\xi^{ex}_l({\bf k},t)\xi^{ex}_l({\bf k}',t')\rangle &=&
\langle\xi^{ex}_\perp({\bf k},t)\xi^{ex}_\perp({\bf k}',t')\rangle =
\frac{1}{n}\frac{2 \gamma_bT_b}{m}  \delta(t-t')\delta({\bf k}+{\bf k}'),
\label{fh.3}
\end{eqnarray}
while the variances of the internal contributions are obtained by
imposing the fluctuation-dissipation theorem (see~\ref{app}
for details):
\begin{eqnarray}
\langle\theta^{in}({\bf k},t)\theta^{in}({\bf k}',t')\rangle &=& 2\kappa T_g^2
\delta(t-t')\delta({\bf k}+{\bf k}') \nonumber  \\
\langle\xi^{in}_l({\bf k},t)\xi^{in}_l({\bf k}',t')\rangle &=&
2 nm\nu_l T_g  \delta(t-t')\delta({\bf k}+{\bf k}') \nonumber \\
\langle\xi^{in}_\perp({\bf k},t)\xi^{in}_\perp({\bf k}',t')\rangle &=&
2 nm\nu T_g  \delta(t-t')\delta({\bf k}+{\bf k}').
\label{fh.4}
\end{eqnarray}
The internal and external noises are uncorrelated

\begin{equation}
\langle\theta^{ex}({\bf k},t)\theta^{in}({\bf k}',t')\rangle = 
\langle\xi_l^{ex}({\bf k},t)\xi_l^{in}({\bf k}',t')\rangle 
= \langle\xi_\perp^{ex}({\bf k},t)\xi_\perp^{in}({\bf k}',t')\rangle = 0.
\label{fh.5}
\end{equation}

The hydrodynamic analysis of model~(\ref{m.2}) consists, then, in
solving the system of coupled linear Langevin equations~(\ref{fh.2}).
In particular we are interested in finding the explicit forms of the
static and dynamical structure factors, respectively

\begin{equation}
S_{ab}({\bf k})=\lim_{t\to\infty}\frac{1}{V}\langle \delta
a({\bf k},t)\delta b({-\bf k},t)\rangle,
\end{equation}
and 

\begin{equation}
S_{ab}({\bf k},\omega)=\int_{-\infty}^{\infty}S_{ab}({\bf k},t)e^{-i\omega t},
\end{equation}
where

\begin{equation}
S_{ab}({\bf k},t)= \lim_{t'\to\infty}\frac{1}{V}
\langle \delta a({\bf k},t'+t)\delta b({-\bf k},t')\rangle.
\end{equation}

\section{Out-of-equilibrium correlations: Static and Dynamical structure factors}
\label{Out-of-equilibrium}

It is well known that spatially extended correlations develop in the
non-equilibrium stationary state of a driven granular
fluid~\cite{NETP99,VAZ11}.  In particular, the velocity correlator
$\langle {\bf u}({\bf k})\cdot{\bf u}(-{\bf k}) \rangle$, where the
average $\langle\ldots \rangle$ is taken over noises, can be written
as
\begin{equation}
\langle {\bf u}({\bf k})\cdot{\bf u}(-{\bf k}) \rangle = \langle u_\parallel({\bf k})u_\parallel(-{\bf k}) \rangle + 
\langle u_\perp({\bf k})u_\perp(-{\bf k}) \rangle,
\label{correlators}
\end{equation}
with the cross terms $\langle u_\perp({\bf k})u_\parallel(-{\bf k})
\rangle =0 $.  The two terms on the right of Eq.~(\ref{correlators})
can be studied separately.

\begin{figure}[tbp!]
\begin{center}
\includegraphics[width=0.6\columnwidth,clip=true]{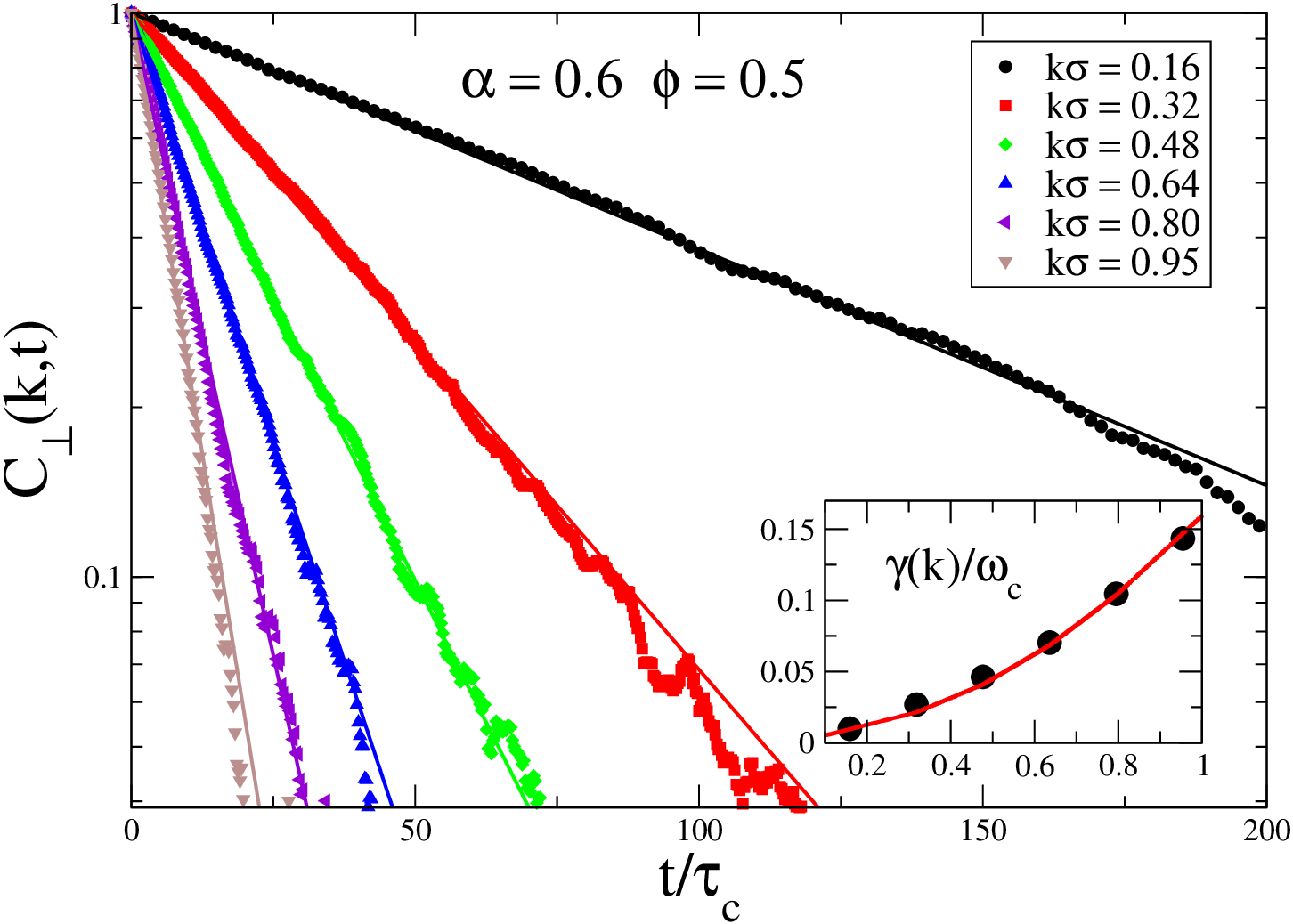}
\caption{$C_\perp(k,t)$ for $\alpha=0.6$ and $\phi=0.5$ and
  several values of $k\sigma$ measured in numerical simulations,
  together with exponential fits (continuous curves). In the inset the
  quantity $\gamma(k)$ obtained from the fits is reported, rescaled
  with the collision frequency $\omega_c$, together with the parabolic
  fit via the formula $\gamma(k)=\gamma_b+\nu k^2$ (continuous red
  line).}
\label{Clt}
\end{center}
\end{figure}

\subsection{Shear modes}
\label{ShearModes}

According to the matrix~(\ref{fh.1}) the shear modes are decoupled
from the others in the linear approximation and their dynamics obeys a
simple Langevin equation
\begin{equation}
\dot{u}_\perp({\bf k},t) = -(\gamma_b + \nu k^2)u_\perp({\bf k},t)+ \xi_\perp^{ex}({\bf k},t)+ik/n~\xi_\perp^{in}({\bf k},t),
\label{langeqSHEAR}
\end{equation}
where, from now on, we put $m=1$ in all
formulas.  The effect of internal and external noises is equal to a
single complex noise $\xi_\perp({\bf k},t) = \xi_\perp^{ex}({\bf
  k},t)+ik/n~\xi_\perp^{in}({\bf k},t)$ with variance:
\begin{equation}
V^{-1}\langle \xi_\perp({\bf k},t) \xi_\perp(-{\bf k},t^{\prime}) \rangle
~=~ \frac{2}{n} \left( T_b\gamma_b + \nu T_g k^2\right) \delta(t-t^{\prime}).
\end{equation}
From this, the \emph{equal-time} correlator $\langle u_\perp({\bf
  k})u_\perp(-{\bf k}) \rangle$ can be easily calculated:
\begin{equation}
nS_\perp(k)=N^{-1}\langle |u_\perp(k)|^2\rangle = 
\frac{\gamma_b T_b + \nu k^2 T_g}{\gamma_b + \nu k^2} = T_g + \frac{ (T_b - T_g)}{ 1 + \xi^2 k^2 },
\label{eq:Sklowk} 
\end{equation}  
with $\xi^2 = \nu/\gamma_b$. The meaning of the above equation is
clear: the small length-scale physics, which depends on the inelastic
collisions, is related to the granular temperature $T_g$, while at
large distances there are correlations with finite amplitude $T_b -
T_g$ and extent $\xi$. Notice that in the limit
  $\gamma_b\to 0$, keeping $\gamma_bT_b$ finite, from
  Eq.~(\ref{eq:Sklowk}) one obtains the result of Ref.~\cite{NETP99},
  where $S_\perp(k)\sim 1/k^2$ and long range correlations are
  observed. A similar power law behaviour is also observed in molecular
  fluids under shear~\cite{SOZ10}. The different result we obtain in
  our model is due to the intrinsic cut-off introduced by the viscous
  drag $\gamma_b>0$.  Indeed, the finite extent of correlations is
even more clear when Eq.~(\ref{eq:Sklowk}) is written in real space,
yielding the spatial correlation function $G_\perp({\bf r})$, which
reads:
\begin{equation}
nG_\perp({\bf r}) 
= T_g \delta^{(2)}({\bf r}) + (T_b - T_g)\frac{K_0(r/\xi)}{\xi^2},
\label{eq:Sx} 
\end{equation} 
where $K_0(x)$ is the 2nd kind modified Bessel function that, for
large distances, decays exponentially
\begin{equation}
K_0(r/\xi)\approx \sqrt{\frac{\pi}{2}}\frac{e^{-r/\xi}}{(r/\xi)^{1/2}}.
\label{eq:Sx2} 
\end{equation}   
At equilibrium, namely when collisions are elastic and $T_g=T_b$,
equipartition between modes is perfectly fulfilled and the structure
factor becomes flat, i.e. $S_\perp(k)=T_b$.  Differently, in the
granular case, where $T_g \ne T_b$, equipartition breaks down and from
Eq.~(\ref{eq:Sklowk}) we have that $S_\perp \to T_b$ for small $k$ and
$S_\perp \to T_g$ for large $k$.  We see here that out of equilibrium
the quantity $\xi=\sqrt{\nu/\gamma_b}$ measures the range of
\emph{static} correlations of the vorticity field. The
  behaviour described above is in good agreement with experimental
  results obtained for driven granular fluids, as reported
  in~\cite{PEU02} and, more recently, in~\cite{GSVP11}.

From Eq.~(\ref{langeqSHEAR}) we also find that fluctuations decay
exponentially
\begin{equation} 
\langle u_\perp(k,t) u_\perp(-k,0)\rangle \sim S_\perp(k)~e^{ -(\gamma_b + \nu k^2)t },
\label{Ctperp}
\end{equation}
with a characteristic time $\tau(k) = \gamma(k)^{-1} = (\gamma_b + \nu
k^2)^{-1}$.  Such a behavior is also observed for elastic fluids, with
the only difference that in that case $S_\perp(k)=T_b/n$ is constant.
The length-scale $\xi = \sqrt{\nu/\gamma_b}$ can be therefore
\emph{always} connected to dynamical properties of the system. What is
peculiar of the out-of-equilibrium regime is that the so defined $\xi$
also represents the extent of correlations of the vorticity field,
thus establishing a remarkable link between static
correlations and dynamical ones.

\begin{figure}[tbp!]
\begin{center}
\includegraphics[width=0.6\columnwidth,clip=true]{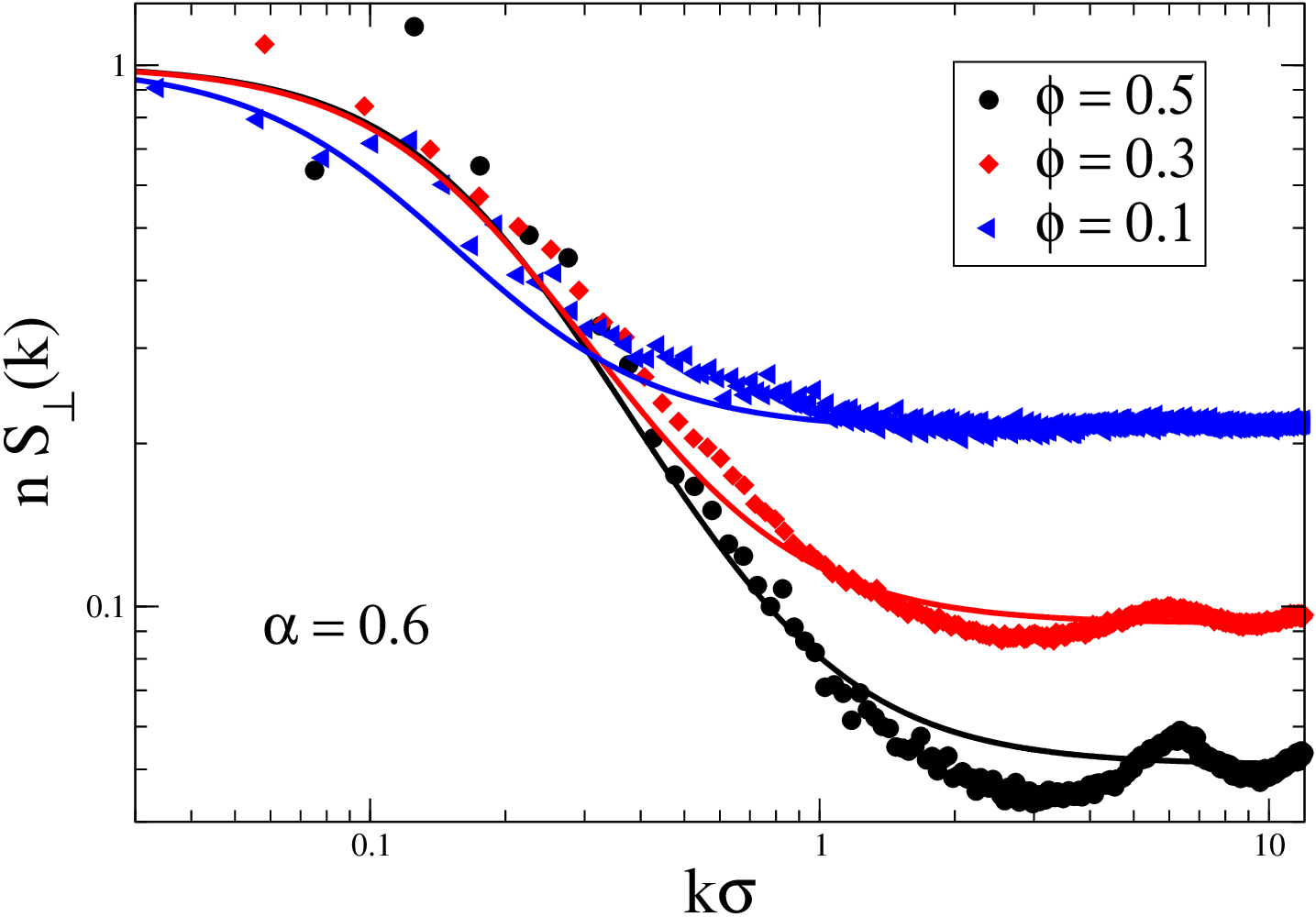}
\caption{Shear modes structure factor $n S_\perp(k)$ at different
  packing fractions for $\alpha=0.6$.  Full curves are drawn by inserting into Eq.~(\ref{eq:Sklowk})
 the values of $\nu$ obtained from the decay of dynamical correlators.}
\label{Sk-trans}
\end{center}
\end{figure}

In Fig.~\ref{Clt} are reported the correlators $C_\perp(k,t)=\langle
u_\perp(k,t) u_\perp(-k,0) \rangle / \langle |u_\perp(k)|^2\rangle$
for different values of $k$ and the same packing fraction measured in
numerical simulations.  Let us notice that the decay of $\langle
u_\perp(k,t) u_\perp(-k,0)\rangle$ is always exponential: we can
therefore \emph{a-posteriori} support the validity of linear
hydrodynamics, from which the Langevin equations for shear modes is
obtained.  By interpolating with a parabola the characteristic time
$\tau(k)$ as a function of $k$, see inset of Fig.~(\ref{Clt}), we
obtain the shear viscosity $\nu$.  Let us stress the deep connection
between statics and dynamics in the out-of-equilibrium regime:
inserting the values of $\nu$ obtained from the dynamics into 
Eq.~(\ref{eq:Sklowk}), we find curves that well superimpose the
numerical data for $\langle u_\perp(k) u_\perp(-k)\rangle$, (see
Fig.~\ref{Sk-trans}).  The values of $\nu$ obtained from the dynamics
of shear modes can be independently obtained as fit parameters of
$S_\perp(k)$ via Eq.~(\ref{eq:Sklowk}).  The values of $\nu$ obtained
with the two different procedures are compatible, as can be seen from
Tab.~\ref{tab:SkPerpNu}.  They are also reasonably close to the dense EHS
predictions, presented in the same Table.

\begin{table}
\begin{center}
\begin{tabular}{|c|c|c|c|}
\hline
$\phi$ & Statics & Dynamics & dense EHS \\ 
\hline
0.1  & 0.0062 & 0.0072 & 0.0111 \\
0.2  & 0.0036 & 0.0041 & 0.0052 \\
0.3  & 0.0025 & 0.0031 & 0.0039 \\
0.4  & 0.0026 & 0.0027 & 0.0037 \\
0.5  & 0.0020 & 0.0028 & 0.0041 \\
\hline
\end{tabular}
\caption{Comparison of the shear viscosity obtained from statics (fit
  via Eq.~(\ref{eq:Sklowk})), dynamics (fit via Eq.~(\ref{Ctperp}))
  and dense EHS approximation.}
\label{tab:SkPerpNu} 
\end{center}
\end{table}

\subsection{Longitudinal modes}
\label{SoundModes}

\subsubsection{Static correlations}
                                   
The same considerations discussed above for shear modes also hold for
the other hydrodynamic modes, which are coupled each other.  In order
to study their behaviour we have to take into account all the
elements of the dynamical matrix.  In particular, the matrix of static
structure factors ${\bf S}({\bf k})$ with elements $S_{a b}(k)$, is
obtained solving the following linear system:
\begin{equation}
{\bf M}(k){\bf S}(k)+{\bf S}^\dagger(k){\bf M}^\dagger(k)+{\bf C}(k) = 0, 
\label{ssf.1}
\end{equation}
where the matrix of noises ${\bf C}(k)$ is such that:
\begin{equation}
V^{-1}\langle {\bf f}({\bf k},t)\otimes{\bf f}(-{\bf k},t')\rangle
= {\bf C}({\bf k}) \delta(t-t^{\prime}).
\end{equation}
Here
\begin{equation}
\hspace{-2.5cm}{\bf C}({\bf k})=\textrm{diag}\left[0,
  \frac{2T_g}{dn}\left(4T_b\gamma_b+2D_T T_g k^2\right),
  \frac{2}{n}\left(T_b \gamma_b+ \nu_l T_g
  k^2\right), \frac{2}{n}\left(T_b\gamma_b + \nu T_g
  k^2\right)\right],
\label{ssf.2}
\end{equation}
where $\textrm{diag}[x,y,z,w]$ denotes a diagonal matrix with elements $x,y,z,w$.

\begin{figure}[tbp!]
\begin{center}
\includegraphics[width=0.6\columnwidth,clip=true]{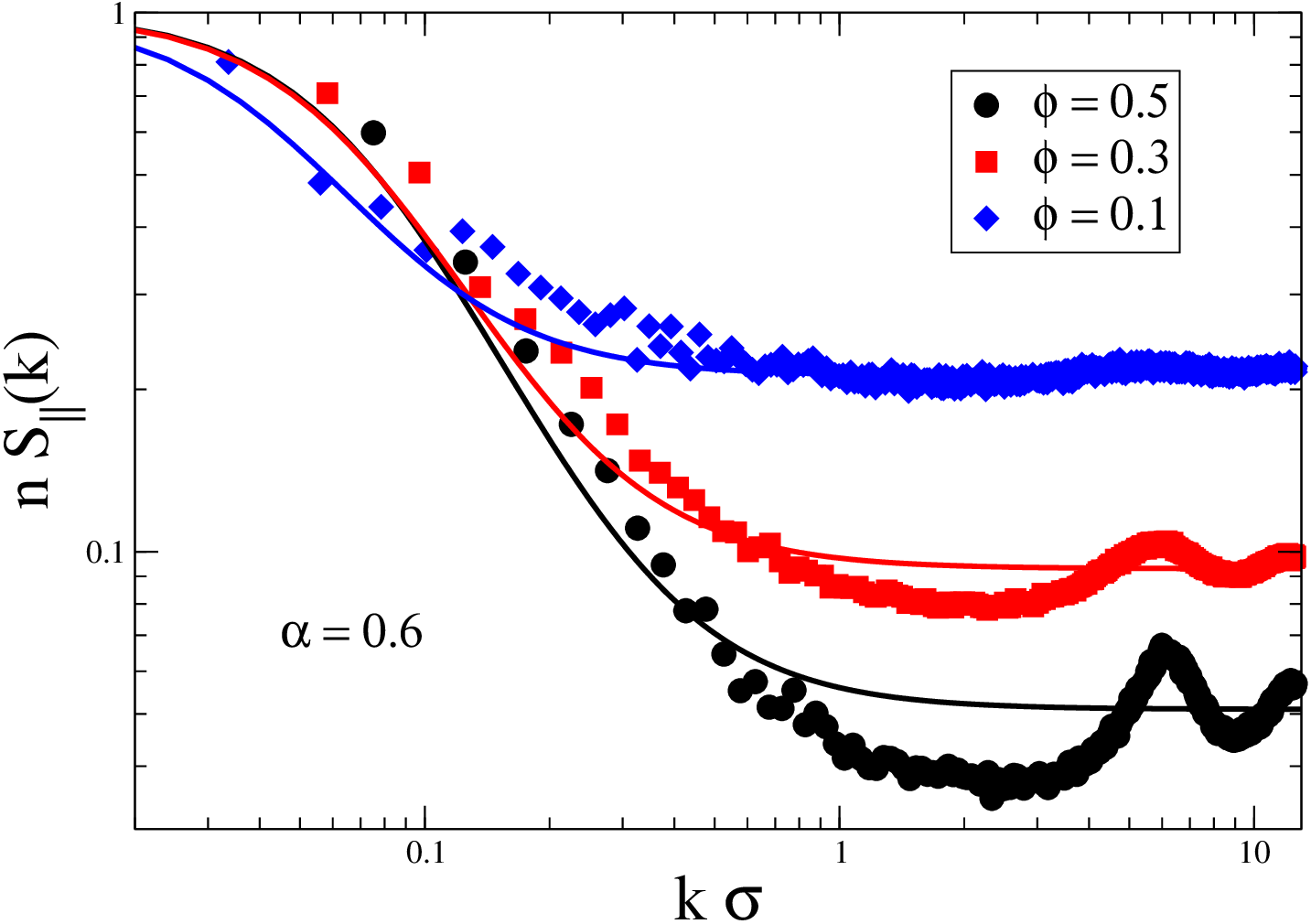}
\caption{Longitudinal modes structure factor $n S_\parallel(k)$ at
  different packing fractions, for $\alpha=0.6$.  Full curves are
  drawn inserting into Eq.~(\ref{eq:longitudinalSK}) the values of
  $\xi_l$, defined in Eq.~(\ref{renormalizedNU}), obtained from best
  fits of the dynamical structure factors.}
\label{Sk-longitudinal}
\end{center}
\end{figure}

The expression of the longitudinal structure factor,
\begin{equation}
n S_{\parallel}(k) = N^{-1} \langle |u_l(k)|^2 \rangle,
\end{equation}
turns out to be the ratio between two even polynomial functions of the $6-$th order 
in $k$:
\begin{equation}
n S_{\parallel}(k)=\frac{S_0 + S_2 k^2 +S_4 k^4 +S_6 k^6}{S_0' + S_2' k^2 +S_4' k^4 +S_6' k^6}.
\label{eq:Skll}
\end{equation} 
In the above expression eight constants have been introduced, which
depend in a complicate manner by all the parameters of the system.
Let us focus here on the asymptotic behavior of $S_{\parallel}(k)$ at
large and small values of $k$. From
  Eq.~(\ref{tg}) there follows the relation
  $d\gamma_0\omega_c/2=\gamma_b(T_b-T_g)/T_g$, which allows us to
  recast the series expansion around $k=0$ of the expression in
  Eq.~(\ref{eq:Skll}) in the form
\begin{equation}
n S_{\parallel}(k\rightarrow 0) \simeq T_b - ( T_b - T_g) \xi_l^2 k^2 + \mathcal{O}(k^2),
\label{Sll-smallk}
\end{equation}
with
\begin{equation}
\hspace{-0.5cm}\xi_l^2 = \frac{\nu^*_l}{\gamma_b}=\frac{1}{\gamma_b}\left[\nu_l + \frac{\gamma_bT_b}{n
    T_g(\gamma_b+\gamma_0\omega_c)(2\gamma_b+3\gamma_0\omega_c)}
  \left(\frac{4p^2}{d^2nT_g}+ \frac{2g(n)p}{3d} \right)\right].
\label{renormalizedNU}
\end{equation} 
Up to $\mathcal{O}(k^2)$ the expression in Eq.~(\ref{Sll-smallk}) is equivalent to
\begin{equation}
n S_{\parallel}(k) = T_g + \frac{(T_b-T_g)}{1 + \xi_l^2~k^2},
\label{eq:longitudinalSK}
\end{equation}
namely a form analogous to the structure factor we found for the shear
modes.  In Fig.~\ref{Sk-longitudinal} we show the static structure
factors for different packing fractions.  Again, notice
  that in the limit $\gamma_b\to 0$ with $\gamma_bT_b$ finite, the
  behaviour $S_{\parallel}(k)\sim 1/k^2$ found in~\cite{NETP99} is
  recovered.  Albeit the above expression is in principle only valid
for low $k$ values, it captures also the large $k$ limit, when fine
oscillations are disregarded. Indeed, expanding Eq.~(\ref{eq:Skll})
for large $k$ values we find:
\begin{equation}
n S_{\parallel}(k \rightarrow \infty) \simeq T_g + \frac{( T_b - T_g)}{\xi_l^2 k^2}.
\end{equation}
Such a discussion shows that even for longitudinal modes the viscosity
$\nu_l$ is related to a finite correlation length, measurable from
static velocity correlations when the system is out of equilibrium
with $T_g<T_b$. The behavior of that length when the packing fraction
is increased will be discussed in the last section.

\subsubsection{Dynamical correlations}

Dynamical correlations for longitudinal modes are less simple than
those for the shear mode, since they are given by a superposition of
different (real and imaginary) exponentials.  The dynamical structure
factors are obtained by solving the equation of motion~(\ref{fh.2}),
which, in the frequency domain, reads as
\begin{equation}
\widetilde{{\bf M}}({\bf k},\omega)\delta\tilde{{\bf a}}({\bf k},\omega)=\tilde{{\bf f}}({\bf k},\omega),
\label{dsf.2}
\end{equation}
where

\begin{equation}
\widetilde{{\bf M}}({\bf k},\omega)=i\omega {\bf I}-{\bf M}({\bf k}),
\label{dsf.3}
\end{equation}
with ${\bf I}$ the identity matrix, 

\begin{equation}
\delta \tilde{a}({\bf k},\omega)=\int_{-\infty}^{\infty} dt~\delta a({\bf k},t)e^{-i\omega t},
\end{equation}
and

\begin{equation}
V^{-1}\langle\tilde{{\bf f}}({\bf k},\omega)\otimes\tilde{{\bf f}}(-{\bf k},\omega')\rangle
={\bf C}({\bf k})\delta(\omega+\omega'),
\label{dsf.4}
\end{equation}
${\bf C}(k)$ being defined in Eq.~(\ref{ssf.2}).  Multiplying
Eq.~(\ref{dsf.2}) on the left by $\widetilde{{\bf M}}^{-1}(k,\omega)$
and on the right by $\delta\tilde{{\bf a}}^{T}(-k,-\omega)$ (where
$X^T$ denotes the transpose of $X$) and averaging over the noise, we
obtain the matrix of dynamical structure factors
\begin{equation}
\hspace{-2cm}{\bf S}(k,\omega)=V^{-1}
\langle\widetilde{{\bf M}}^{-1}(k,\omega)\tilde{{\bf f}}(k)\delta\tilde{{\bf a}}^{T}(-k,-\omega)\rangle 
=\widetilde{{\bf M}}^{-1}(k,\omega){\bf C}(k)
[\widetilde{{\bf M}}^{T}(-k,-\omega)]^{-1},
\label{dsf.5}
\end{equation}
where in the last equality we have used the Hermitian conjugate of
Eq.~(\ref{dsf.2}) and the relation~(\ref{dsf.4}).

The dynamical structure factors $S_{nn}(k,\omega)$ and $S_{\parallel}(k,\omega)$ take the explicit forms 
\begin{eqnarray}
\label{SKrhorho}
\hspace{-2.5cm}S_{nn}(k,\omega)=n^2k^2\left(\frac{\left[\omega^2+(2\gamma_b+D_Tk^2+3\gamma_0\omega_c)^2\right]
(\frac{2\gamma_bT_b}{n}+\frac{2\nu k^2T_g}{n})
+k^2\left(\frac{p}{n T_g}\right)^2\left(\frac{4D_TT_g^2k^2}{nd}+\frac{8T_g\gamma_bT_b}{nd}\right)}
{|\det \widetilde{{\bf M}}|^2}\right), \nonumber \\
\end{eqnarray}
where

\begin{eqnarray}
\label{determinante}
&&\hspace{-2.5cm}|\det \widetilde{ {\bf M}}|^2 = \left[-\omega^2(3\gamma_0\omega_c+D_Tk^2+\nu
k^2+3\gamma_b)+k^2\left((2\gamma_b+3\gamma_0\omega_c)v_T^2
-\frac{g(n)p\gamma_0\omega_c}{n}+v_T^2D_Tk^2\right)\right]^2
\nonumber \\ 
&&\hspace{-2cm}+\left\{\omega^3-\omega\left[\gamma_b(2\gamma_b+3\gamma_0\omega_c)
+k^2\left(D_T(\nu k^2+\gamma_b)
+\nu(2\gamma_b+3\gamma_0\omega_c)+\frac{2p^2}{dn^2mT_g}+v_T^2\right)\right]\right\}^2,
\nonumber \\
\end{eqnarray}

and
\begin{equation}
S_{\parallel}(k,\omega)=\frac{\omega^2}{n^2k^2}S_{nn}(k,\omega).
\label{dfs.6}
\end{equation}

\begin{table}[htb!]
\centering
\begin{tabular}{|r||c|c||c|c|} \hline \hline
\multicolumn{5}{|c|}{$\phi=0.5$ } \\ \hline 
 & \multicolumn{2}{|c|} {$\alpha=0.6$} & \multicolumn{2}{|c|} {$\alpha=0.8$} \\ \hline 
 & \multicolumn{1}{|c|} {Sim} & \multicolumn{1}{|c|} {dense EHS} & \multicolumn{1}{|c|} {Sim} & \multicolumn{1}{|c|} {dense EHS} \\ \hline 
  $T_g$                     & 0.051 & 0.0416    &  0.066 &  0.0603   \\ \hline
  $\omega_c$                & 179   & 160       &  181   & 181       \\ \hline \hline
\multicolumn{5}{|c|}{$\phi=0.3$ } \\ \hline 
 & \multicolumn{2}{|c|} {$\alpha=0.6$} & \multicolumn{2}{|c|} {$\alpha=0.8$} \\ \hline 
 & \multicolumn{1}{|c|} {Sim} & \multicolumn{1}{|c|} {dense EHS} & \multicolumn{1}{|c|} {Sim} & \multicolumn{1}{|c|} {dense EHS} \\ \hline 
  $T_g$                     & 0.093 & 0.0829     & 0.125 &  0.1185   \\ \hline
  $\omega_c$                & 79    & 83         & 85    &  85       \\ \hline \hline
\multicolumn{5}{|c|}{$\phi=0.1$ } \\ \hline 
 & \multicolumn{2}{|c|} {$\alpha=0.6$} & \multicolumn{2}{|c|} {$\alpha=0.8$} \\ \hline 
 & \multicolumn{1}{|c|} {Sim} & \multicolumn{1}{|c|} {dense EHS} & \multicolumn{1}{|c|} {Sim} & \multicolumn{1}{|c|} {dense EHS} \\ \hline 
  $T_g$                     & 0.212 & 0.2055      & 0.286 & 0.2820   \\ \hline
  $\omega_c$                & 26    & 25          & 29    & 28       \\ \hline \hline
\end{tabular}
\caption{Comparison of theoretical predictions of Eqs.~(\ref{tg})
  and~(\ref{omegac}) and numerical results for $T_g$ and $\omega_c$.}
\label{Tgtable}
\end{table}

All the coefficients appearing in these expressions can be evaluated
within the dense EHS approximation, so that Eq.~(\ref{dfs.6}) can be used
in order to obtain $\nu_l$ and $D_T$ from the numerical data. 
Clearly the writing of physical quantities, such as $v_T^2=(\partial p/\partial n)_T$
or $p$ itself, with dense EHS formulas leads to systematic errors.  The
amplitude of this error can be estimated for instance by comparing
the theoretical prediction of dense EHS collision frequency with the
collision frequency measured in simulations.  In Table \ref{Tgtable}
are listed both dense EHS and numerical collision frequencies at
different packing fractions. Again it is found that the dense EHS
prediction is quite good.  

\begin{figure}[tbp!]
\begin{center}
\includegraphics[width=0.6\columnwidth,clip=true]{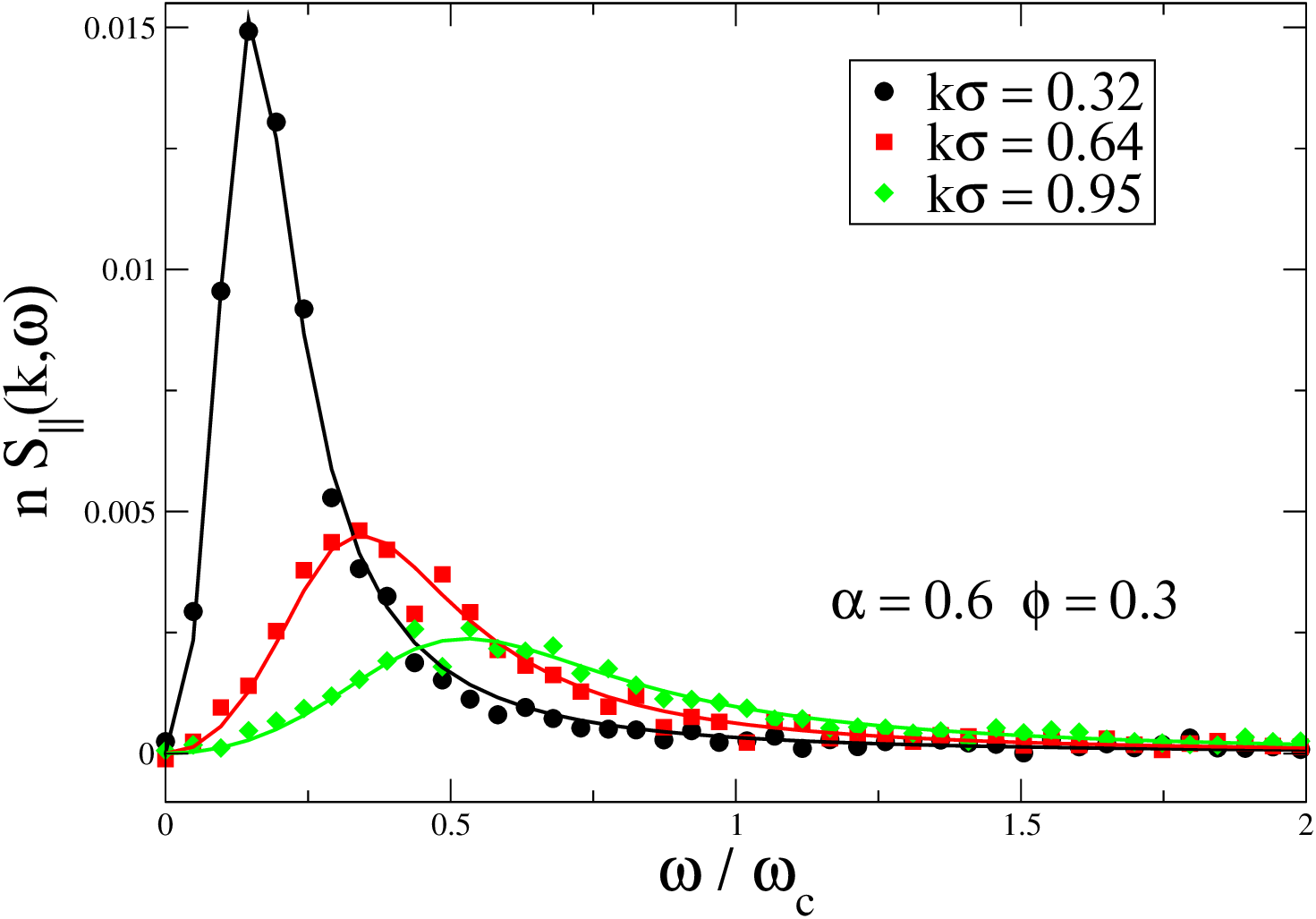}
\caption{Dynamical structure factor $n S_\parallel(k,\omega)$ at fixed
  packing fractions $\phi$ and different momenta. Identifying a
  characteristic time $\tau_{peak}$ with the frequency of the maximum
  of the curve we see that $\tau_c/\tau_{peak}=\omega/\omega_c$ grows
  with the wave-vector, namely longitudinal modes of
  higher momentum decay faster.  Continuous lines show the best fit
  results via Eq.~(\ref{dfs.6}), with $D_T$ and $\nu_l$ as fitting
  parameters and all other coefficients fixed with the dense EHS
  predictions.}
\label{FitDinamici}
\end{center}
\end{figure}

In order to obtain the longitudinal viscosity and the thermal
diffusion coefficient, we fit our numerical data for the longitudinal
modes using Eq.~(\ref{dfs.6}), where all parameters but $\nu_l$ and
$D_T$ are fixed to the dense EHS values and $T_g$ is the one
measured in simulations.  In Fig.~\ref{FitDinamici}
$S_{\parallel}(k,\omega)$ is shown for different values of $k$ and
fixed $\phi=0.5$, together with the best fit curves.  The values of
$\nu_l$ and $D_T$ so obtained, together with those computed within the
dense EHS approximation, for different values of $\phi$ and $\alpha$, are
reported in Table~\ref{DTnultable} and within errors are found
independent of $k$.  This fact represents an \emph{a-posteriori} check
that we are in the regime of validity of linearized
hydrodynamics. Indeed, in Eqs.~(\ref{dfs.6})
and~(\ref{determinante}) $\nu_l$ appears as $k$-independent variable.
Only at low packing fraction we observe a dependence on $k$.  This is
perhaps due to diluteness, which implies too large mean free path or
mean free time with respect to mesoscopic scales.

\begin{figure}[tbp!]
\begin{center}
\includegraphics[width=0.6\columnwidth,clip=true]{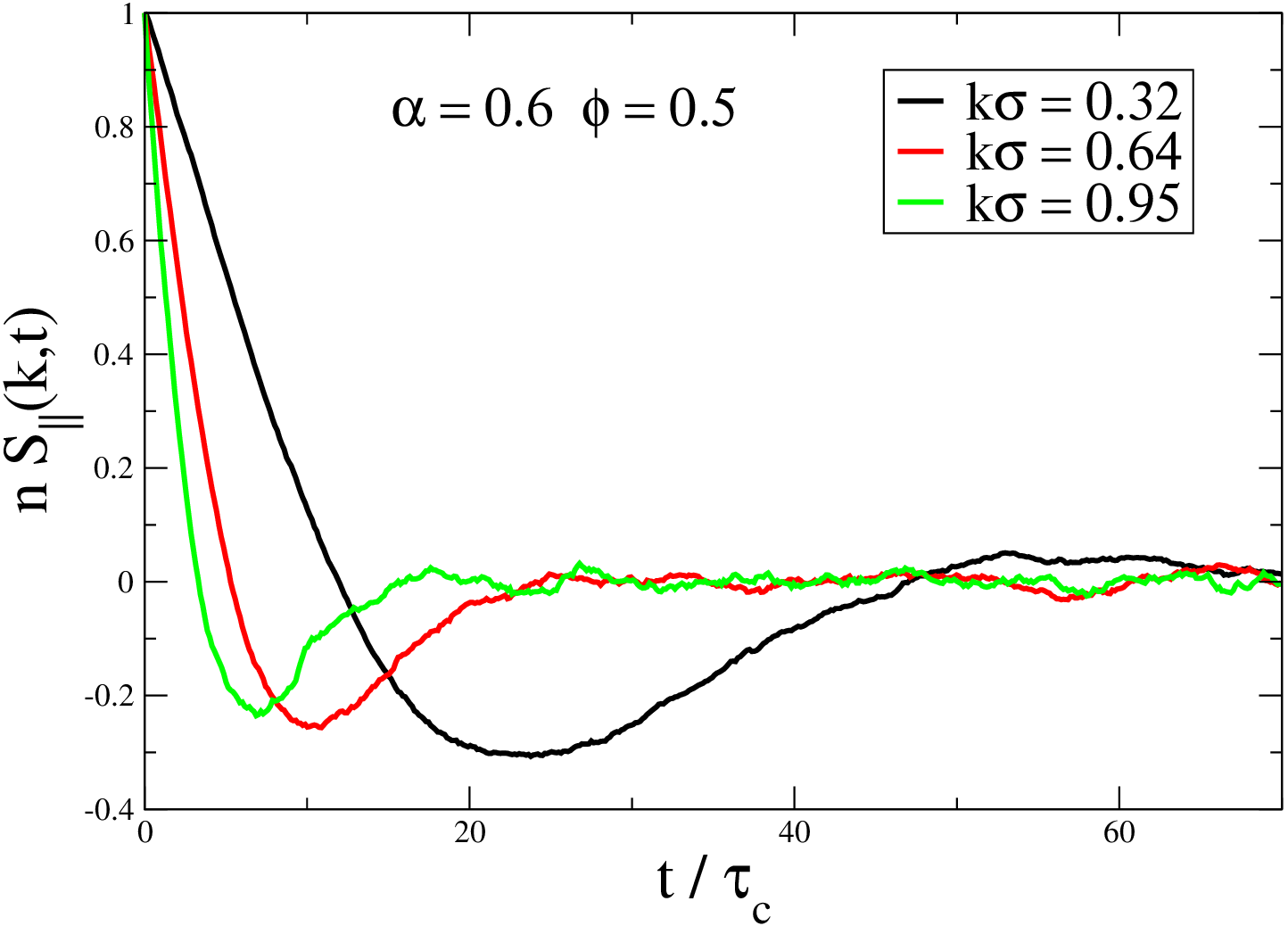}
\caption{Dynamical structure factor $n S_\parallel(k,t)$ at fixed
  packing fraction $\phi$ and different momenta. The observed
  oscillations are in agreement with the eigenvalues spectrum in the
  interval of momenta considered: the eigenvalues of sound modes are
  complex conjugate, thus producing an oscillatory relaxation.}
\label{Sktempo}
\end{center}
\end{figure}

In Fig.~\ref{Sktempo} we also show the time decay of the dynamical
structure factor.  It can be appreciated, in the time domain, the
superposition of different real and imaginary exponentials, which
determines a mix of damping and propagation.

\begin{figure}[tbp!]
\begin{center}
\includegraphics[width=0.6\columnwidth,clip=true]{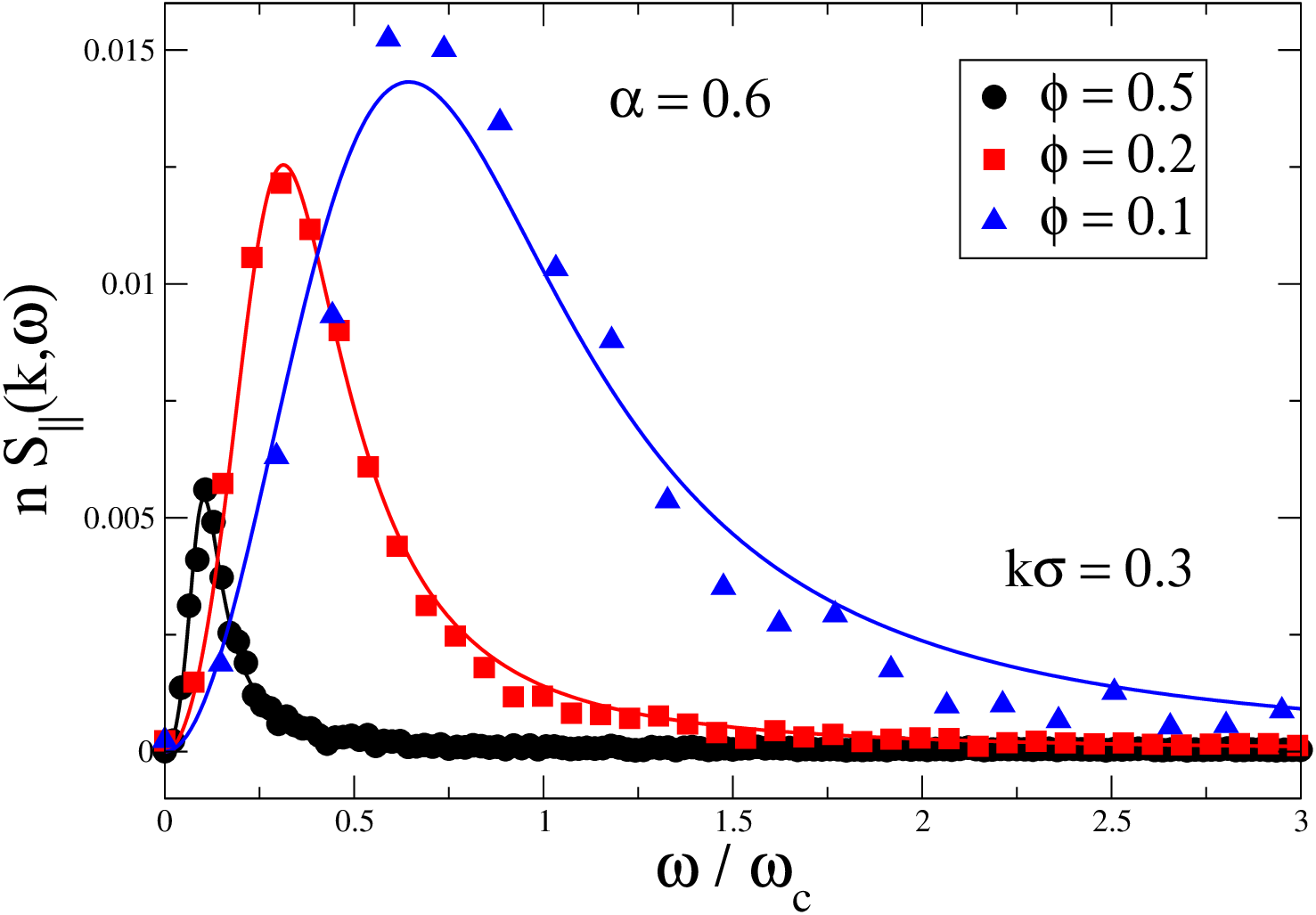}
\caption{Dynamical structure factor $n S_\parallel(k,\omega)$ at fixed
  momentum $k$ and different packing fractions. Identifying a
  characteristic time $\tau_{peak}$ with the frequency of the maximum
  of the curve we see that $\tau_c/\tau_{peak}=\omega/\omega_c$ grows
  when the packing fraction is lowered, namely sound modes decays
  faster compared to the microscopic time-scale. Continuous lines show
  the best fit results via Eq.~(\ref{dfs.6}), with $D_T$ and $\nu_l$
  as fitting parameters and all other coefficients fixed with the
  dense EHS predictions.}
\label{Skfrequenza}
\end{center}
\end{figure}

The dynamical structure factor at fixed $k$ and different packing
fractions is reported in Fig.~\ref{Skfrequenza}: it is remarkable
the observation of a time-scale, individuated by the peak frequency of
$S_{\parallel}(k,\omega)$, which increases as the packing fraction is
increased. Such behaviour is consistent with the
observation, discussed in details below, of a {\em growth} of the
correlation lengths defined above, together with the packing fraction.

\begin{table}[htb]
\centering
\begin{tabular}{|r||c|c||c|c|} \hline \hline
\multicolumn{5}{|c|}{$\phi=0.5$ } \\ \hline 
 & \multicolumn{2}{|c|} {$\alpha=0.6$} & \multicolumn{2}{|c|} {$\alpha=0.8$} \\ \hline 
  & $D_T$ & $\nu_l$ & $D_T$ & $\nu_l$ \\ \hline 
Fit Results: $k\sigma=0.5$ & 0.019  &  0.0053 & 0.020 & 0.011 \\ \hline
             $k\sigma=0.6$ & 0.020  &  0.0046 & 0.018 & 0.011 \\ \hline
             $k\sigma=0.8$ & 0.021  &  0.0055 & 0.019 & 0.0076 \\ \hline
dense EHS & 0.018 & 0.0081 & 0.021 & 0.0090 \\ \hline \hline 
\multicolumn{5}{|c|}{$\phi=0.3$ } \\ \hline 
 & \multicolumn{2}{|c|} {$\alpha=0.6$} & \multicolumn{2}{|c|} {$\alpha=0.8$} \\ \hline 
  & $D_T$ & $\nu_{\rm l}$ & $D_T$ & $\nu_{\rm l}$ \\ \hline 
Fit Results:  $k\sigma=0.4$ & 0.017 &  0.0052  & 0.013  & 0.0098 \\ \hline
              $k\sigma=0.5$ & 0.020 &  0.0058  & 0.013  & 0.0091 \\ \hline
              $k\sigma=0.6$ & 0.017 &  0.0058  & 0.015  & 0.0079 \\ \hline
dense EHS & 0.018 & 0.0057 & 0.021 & 0.0066 \\ \hline \hline 
\multicolumn{5}{|c|}{$\phi=0.1$ } \\ \hline 
 & \multicolumn{2}{|c|} {$\alpha=0.6$} & \multicolumn{2}{|c|} {$\alpha=0.8$} \\ \hline 
  & $D_T$ & $\nu_{\rm l}$ & $D_T$ & $\nu_{\rm l}$ \\ \hline 
Fit Results:  $k\sigma=0.2$ & 0.039 & 0.016  & 0.021  & 0.023 \\ \hline
              $k\sigma=0.3$ & 0.028 & 0.016  & 0.019  & 0.018 \\ \hline
              $k\sigma=0.4$ & 0.018 & 0.013  & 0.0096  & 0.016 \\ \hline
dense EHS  & 0.048 & 0.012 & 0.056 & 0.014 \\ \hline \hline 
\end{tabular}
\caption{Comparison of theoretical predictions of
  Eqs.~(\ref{nuenskog}-\ref{kappa-enskog}) and fit results via
  Eqs.~(\ref{dfs.6}) for $D_T$ and $\nu_l$.}
\label{DTnultable}
\end{table}

 We conclude this section by stressing the remarkable agreement between
 numerical $S_{\parallel}(k)$ data and the expression of
 Eq.~(\ref{eq:longitudinalSK}), see Fig.~\ref{Sk-longitudinal}, with $\nu_l$ measured from dynamics
 and the renormalization term entering the definition of $\nu_l^*$
 (see Eq.~(\ref{renormalizedNU})) calculated within the dense EHS
 approximation and reported in Table~\ref{DTnultable}.

\section{Summary and conclusions: transport coefficients and non-equilibrium correlation lengths }
\label{Summary}

In conclusion, we have studied both static and dynamical correlations
for hydrodynamic fluctuations of the velocity and density fields:
indeed we recall Eq.~(\ref{dfs.6}) which gives a direct relation
between density and longitudinal velocity structure factors.  A main
comment concerns the good success of analytical predictions:
comparison with simulations shows a fair agreement up to $\phi=0.5$,
with values for most of the parameters directly given in the dense EHS
approximation. This signals a success of the three main ingredients:
1) dense EHS theory for transport coefficients, working even at quite
high densities, 2) scale separation and granular hydrodynamics, and 3)
prescription of Eqs.~(\ref{fh.3}-\ref{fh.5}) for the hydrodynamic
noise. The last ingredient is perhaps the most interesting, if one
considers how hard is the characterization in simple terms of
non-equilibrium systems. Moreover, the assumption of white internal
noise is less obvious than in the case of simple random kicks without
drag~\cite{MGT09}: in that case, the absence of drag ($\gamma_b=0$)
implies a relaxation time at large scales, $k\to 0$, which diverges,
making a solid base for assuming {\em fast} the relaxation of noise
due to microscopic degrees of freedom. In our model, in principle, the
drag could be large enough to make even large scales fast, making
difficult to define hydrodynamic fields. Recent experiments show that,
even if that could be a realistic situation, this does not change
dramatically the qualitative behavior of structure
factors~\cite{GSVP11}.

\begin{figure}[tbp!]
\begin{center}
\includegraphics[width=0.6\columnwidth,clip=true]{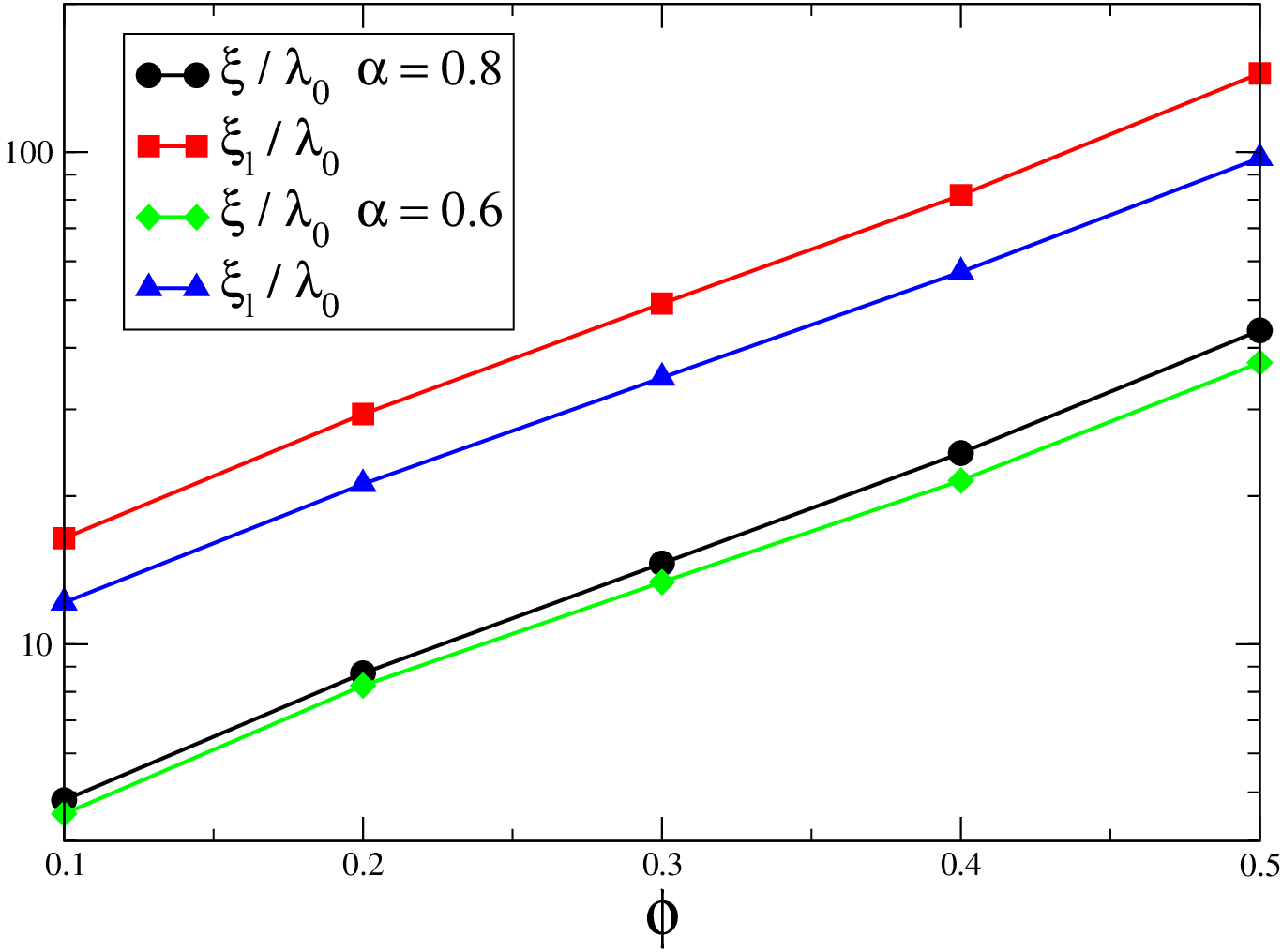}
\caption{Correlation lengths $\xi$ and $\xi_l$ rescaled with the mean
  free path $\lambda_0$, for different packing fractions and several
  values of the restitution coefficient. The extent of
    correlations grows exponentially with the packing fraction.
  Notice also that, at fixed $\phi$, $\xi$ and $\xi_l$ are larger for
  higher values of $\alpha$. However the amplitude of correlations,
  ruled by $T_b-T_g$, is smaller, see
  Eqs.~(\ref{eq:Sklowk}) and~(\ref{Sll-smallk}).}
\label{xi}
\end{center}
\end{figure}

One of the main results of our study is the presence of spatial order
in the form of non-equilibrium velocity correlations.  This should be
related to the slowing-down of the dynamics with increasing packing
fraction in granular systems~\cite{KSZ10,VAZ11}, and with the
existence of a time-scale growing with the density, as observed
in~\cite{SVGP10}.  As discussed in Sec.~\ref{Out-of-equilibrium}, in
our model the competition between different relaxation mechanisms
given by the kinematic and longitudinal viscosities $\nu$ and $\nu_l$
and the thermostat damping $\gamma_b$ give place to a couple of
length-scales characterizing non-equilibrium structure factors:
$\xi=\sqrt{\nu/\gamma_b}$ and $\xi_l=\sqrt{\nu_l^*/\gamma_b}$.  The
first trivial observation is that allowing $\gamma_b \rightarrow 0 $
such lengths diverge.  It means that, according to the prediction of
\cite{NETP99}, the largest-scale correlations are always equal to the
size of the system, which is the largest size available.  In our model
there is a cut-off on such correlations imposed by the viscous drag
$\gamma_b>0$ due to the interaction with the thermal bath. The
existence of such a cut-off, which represents a fixed parameter at
different packing fractions, allows us to show that by increasing the
packing fraction the extent of correlations is \emph{effectively}
increased.  While the absolute value of $\xi$ is only
  slightly enhanced when $\phi$ is increased, as also observed in
  experiments in~\cite{PEU02}, we stress that, in order to appreciate
  the physical meaning of such length at different densities, one has
  to compare it with the microscopic relevant spatial scale in the
  system, which is given by the mean free path of the particles
  $\lambda_0=\sqrt{\pi}\sigma/(8\phi\chi(\phi))$ and which also changes
  with the packing fraction.  Then one finds that $\xi/\lambda_0$ is
  remarkably increased at high densities, as can be seen in
  Fig.~\ref{xi}, and also in recent experiments~\cite{GSVP11}.  We
may summarize the observed phenomenon saying that the higher is the
packing fraction the higher must be the number of intermediate
scattering events between two different particles in order to
decorrelate their velocities.  This scenario could be reflected in
transitions of dynamical origin, at higher packing
fractions~\cite{OU05,RIS06,WT08,KSZ10}.

\section*{Acknowledgments}

We thank U.~M.~B.~Marconi and A.~Vulpiani for helpful suggestions.
The work of the authors is supported by the ``Granular-Chaos''
project, funded by the Italian MIUR under the FIRB-IDEAS grant number
RBID08Z9JE.

\appendix

\section{Dense EHS formulas for hydrodynamic coefficients}
\label{Enskog}

The definitions of the parameters entering the matrix ${\bf M}$ are
\begin{eqnarray}
v_T^2 &=& \left[ \frac{\partial p}{\partial \rho} \right]_T, \\
\omega_c &=& \Omega_d\chi(\phi)n\sigma^{d-1}\sqrt{\frac{T_g}{\pi m}} \label{omegac} \\
g(\phi) &=& 2 \left( 1 + \frac{\phi}{\chi(\phi)}\frac{\partial \chi(\phi)}{\partial \phi} \right).
\end{eqnarray}
In the dense EHS approximation the pressure $p$ can be written as
\begin{equation}
p(n) = nT_g \left( 1 + \frac{\Omega_d\chi n\sigma^d}{2d}\frac{1+\alpha}{2} \right), \nonumber \\
\label{pressure}
\end{equation}
which in $d=2$ reads as
\begin{equation}
p(\phi) = \frac{4}{\pi \sigma^2}\phi~T_g ( 1 + \phi\chi(\phi)(1+\alpha) ),
\end{equation}
where has been made use of the relation between packing fraction and
density $n = 4 \phi / (\pi \sigma^2)$, and of the definition
$\Omega_d=2\pi^{d/2}/\Gamma(d/2)$ with $d=2$. Notice
  also that we have taken into account the correction due to the
  inelasticity, as given in~\cite{PTNE01}. There are dense EHS
formulas also for the dependence of the diffusion coefficients on the
packing fraction and the granular temperature. In the following such
formulas are written for a 2d system. In this case, we use the
Verlet-Levesque approximation for the pair correlation function at
contact: $\chi(\phi)=(1-7\phi/16)/(1-\phi)^2$.

Shear viscosity: 
\begin{equation}
\eta_{E}= \nu_{0} \left[ \frac{1}{\chi(\phi)}+2\phi+
\left( 1+\frac{8}{\pi} \right)\chi(\phi)\phi^2 \right],
\label{nuenskog}
\end{equation}
with
\begin{equation}
\nu_{0} = \frac{1}{2\sigma} \left( \frac{ m T_g}{\pi} \right)^{1/2}.
\end{equation}

Bulk viscosity:
\begin{equation}
\zeta_{E}= \frac{8\phi^2\chi(\phi)}{\pi\sigma} \left( \frac{ m T_g}{\pi} \right)^{1/2}.
\end{equation}

Thermal diffusivity:
\begin{equation}
\kappa_{E}= \kappa_{0} \left[ \frac{1}{\chi(\phi)}+3\phi+
\left(\frac{9}{4}+\frac{4}{\pi}\right) \chi(\phi)\phi^2 \right],
\end{equation}
with
\begin{equation}
\kappa_{0} = \frac{2 }{\sigma} \left( \frac{ T_g}{\pi m} \right)^{1/2}.
\label{kappa-enskog}
\end{equation}

\section{Noises}
\label{app}

In this appendix we present a detailed discussion of the noise terms
appearing in the fluctuating hydrodynamic equations~(\ref{fh.2}).  Let
us start from the \emph{external} noises, which can be simply obtained
from Eqs.~(\ref{m.2}) and~(\ref{he.0}). We find that the noise
contributions to the equations for the velocity and temperature fields
are, respectively

\begin{eqnarray}
\boldsymbol{\xi}^{ex}({\bf r},t)&=&\frac{1}{n}\sum_i \boldsymbol{\xi}_{b,i}(t)\delta({\bf r}-{\bf r}_i(t)) \label{a.0} \\
\theta^{ex}({\bf r},t)&=&\frac{2m}{dn}\sum_i {\bf v}_i(t)\cdot\boldsymbol{\xi}_{b,i}(t)\delta({\bf r}-{\bf r}_i(t)).
\label{a.1}
\end{eqnarray}
These are Gaussian noises with variances

\begin{eqnarray}
\langle\xi^{ex}_\alpha({\bf r},t)\xi^{ex}_\beta({\bf r'},t')\rangle
&=&\frac{1}{n}\frac{2\gamma_bT_b}{m} \delta_{\alpha\beta}\delta(t-t')\delta({\bf r}-{\bf r'}) \nonumber \\ 
\langle\theta^{ex}({\bf r},t)\theta^{ex}({\bf r'},t')\rangle &=& 
\frac{4mT_g}{dn}\frac{2\gamma_bT_b}{m}\delta(t-t')\delta({\bf r}-{\bf r'}).
\label{a.2}
\end{eqnarray}

In order to describe the local \emph{spontaneous} microscopic
fluctuations of the fluid, we also include in our description
internal \emph{conserved} Gaussian noises $\boldsymbol{\theta}^{in}$ and
$\boldsymbol{\xi}^{in}$, which enter the constitutive equations for
${\bf J}$ and $\boldsymbol{\Pi}$, respectively
 
\begin{eqnarray} 
{\bf J}&=&-\kappa\boldsymbol{\nabla}T + \boldsymbol{\theta}^{in}  \label{a.3} \\
\boldsymbol{\Pi}&=&p{\bf 1}-\eta\left[\boldsymbol{\nabla}{\bf u}+
\left(\boldsymbol{\nabla}{\bf u}\right)^{\dag}\right]+
\left(\frac{2}{d}\eta-\zeta\right){\bf 1}\boldsymbol{\nabla}\cdot{\bf u} + \boldsymbol{\xi}^{in},
\label{a.4}
\end{eqnarray}
where $\kappa$ is the heat conductivity, $p$ is the local pressure,
${\bf 1}$ the unit tensor, $\eta$ the shear viscosity and $\zeta$ the bulk
viscosity.  The amplitudes of such noises are obtained from the
fluctuation-dissipation theorem~\cite{LandauFisStat,FU70,BPRV08}

\begin{eqnarray}
\langle\xi^{in}_{\alpha\beta}({\bf r},t)\xi^{in}_{\gamma\delta}({\bf r'},t')\rangle
&=& 2T_g[\eta(\delta_{\alpha\gamma}\delta_{\beta\delta}+\delta_{\alpha\delta}\delta_{\beta\gamma})
+\left(\zeta-\frac{2}{d}\eta\right)\delta_{\alpha\beta}\delta_{\gamma\delta}]
\delta(t-t')\delta({\bf r}-{\bf r'})\nonumber \\ 
\langle\theta^{in}_\alpha({\bf r},t)\theta^{in}_\beta({\bf r'},t')\rangle &=&
2\kappa T_g^2 \delta_{\alpha\beta}\delta(t-t')\delta({\bf r}-{\bf r'}).
\label{a.5}
\end{eqnarray}
Notice that, for granular systems, Eq.~(\ref{a.3}) should be modified,
adding the term $-\mu\nabla n$ on the rhs, which takes into account
the contribution to the heat current due to density
gradients~\cite{JR85}. However, the transport coefficient $\mu$ is
very small for driven systems~\cite{GM02}, and therefore we have
neglected this contribution.

\subsection*{Linearization}

In order to obtain the linear approximation of the hydrodynamic
equations, we consider how a \emph{homogeneous} fluctuation
of the temperature relaxes around its stationary value $T_g$. Therefore we start by
linearizing Eq.~(\ref{he.1}) around $T_g$, with $n({\bf r},t)=n,
{\bf u}=0, T({\bf r},t)=T(t)$, and obtain
\begin{equation}
\delta\dot{T}(t)=-(2\gamma_b/m +3\gamma_0\omega_c)\delta T(t).
\label{a.50}
\end{equation}
Notice that the quantity $-(2\gamma_b/m+3\gamma_0\omega_c)$ coincides with the
heat mode eigenvalue $\lambda_H(k=0)$.

Next, taking the linear terms around the non-equilibrium steady state
in Eqs.~(\ref{he.1}), introducing the external noises~(\ref{a.0})
and~(\ref{a.1}), and using Eqs.~(\ref{a.3}) and~(\ref{a.4}), we obtain

\begin{eqnarray}
\partial_{t}\delta n({\bf r},t) &=& - n \boldsymbol{\nabla}\cdot{\bf
  u}({\bf r},t) \nonumber \\ 
\partial_{t}{\bf u}({\bf r},t) &=& -
\frac{1}{\rho}\boldsymbol{\nabla}p({\bf r},t) + \nu\nabla^{2}{\bf u}({\bf
  r},t) + (\nu_l-\nu)\boldsymbol{\nabla}\boldsymbol{\nabla}\cdot{\bf
  u}({\bf r},t) \nonumber \\
&-& \frac{\gamma_b}{m}{\bf u}({\bf r},t) -\frac{1}{\rho} \boldsymbol{\nabla}\cdot\boldsymbol{\xi}^{in}({\bf r},t) 
  + \boldsymbol{\xi}^{ex}({\bf r},t)  \label{a.5bis} \\ 
\partial_t\delta T({\bf r},t) &=& \frac{2
  \kappa}{nd}\nabla^2\delta T({\bf
  r},t)-\frac{2p}{nd}\boldsymbol{\nabla}\cdot{\bf u}({\bf r},t) 
-\frac{\gamma_0\omega_cg(n)T_g}{n}\delta n\nonumber \\
&-& 
2\frac{\gamma_b}{m} \delta T({\bf r},t) - 3\gamma_0 \omega_c ~\delta T({\bf
  r},t) -\frac{2}{nd} \boldsymbol{\nabla}\cdot\boldsymbol{\theta}^{in}({\bf r},t) + {\theta}^{ex}({\bf r},t), \nonumber
\end{eqnarray}
with $\rho\nu=\eta$ and $\rho\nu_l=2\eta(d-1)/d+\zeta$.  

The velocity field ${\bf u}$ can be split in longitudinal and
transverse components ${\bf u}={\bf u}_l+{\bf u}_\perp$, where ${\bf
  \nabla}\cdot {\bf u}_\perp=0$ and ${\bf \nabla}\wedge {\bf
  u}_\parallel=0$. With such a decomposition Eqs.~(\ref{a.5bis}) can be
written

\begin{eqnarray}
\partial_{t}\delta n({\bf r},t) &=& - n \boldsymbol{\nabla}\cdot{\bf
  u}({\bf r},t) \nonumber \\ 
\partial_{t}{\bf u}_l({\bf r},t) &=& -
\frac{1}{\rho}\boldsymbol{\nabla}p({\bf r},t) + \nu_l\boldsymbol{\nabla}\boldsymbol{\nabla}\cdot{\bf
  u}_l({\bf r},t) - \frac{\gamma_b}{m}{\bf u}_l({\bf r},t) -\frac{1}{\rho} \boldsymbol{\nabla}\cdot\boldsymbol{\xi}_l^{in}({\bf r},t) 
  + \boldsymbol{\xi}_l^{ex}({\bf r},t)  \nonumber \\ 
\partial_{t}{\bf u}_\perp({\bf r},t) &=& \nu\nabla^{2}{\bf u}_\perp({\bf
  r},t) - \frac{\gamma_b}{m}{\bf u}_\perp({\bf r},t) -\frac{1}{\rho} \boldsymbol{\nabla}\cdot\boldsymbol{\xi}_\perp^{in}({\bf r},t) 
  + \boldsymbol{\xi}_\perp^{ex}({\bf r},t)  \label{a.6} \\ 
\partial_t\delta T({\bf r},t) &=& \frac{2
  \kappa}{nd}\nabla^2\delta T({\bf
  r},t)-\frac{2p}{nd}\boldsymbol{\nabla}\cdot{\bf u}({\bf r},t) -
2\frac{\gamma_b}{m} \delta T({\bf r},t) - 3\gamma_0 \omega_c ~\delta T({\bf
  r},t) \nonumber \\
&-&\frac{\gamma_0\omega_cg(n)T_g}{n}\delta n-\frac{2}{nd} \boldsymbol{\nabla}\cdot\boldsymbol{\theta}^{in}({\bf r},t) + {\theta}^{ex}({\bf r},t), \nonumber
\end{eqnarray}
where 

\begin{eqnarray}
\langle\xi^{in}_{\perp,\alpha\beta}({\bf r},t)\xi^{in}_{\perp,\gamma\delta}({\bf r'},t')\rangle
&=& 2T_g\eta(\delta_{\alpha\gamma}\delta_{\beta\delta}+\delta_{\alpha\delta}\delta_{\beta\gamma}
-2\delta_{\alpha\beta}\delta_{\gamma\delta})
\delta(t-t')\delta({\bf r}-{\bf r'})\nonumber \\ 
\langle\xi^{in}_{l,\alpha\beta}({\bf r},t)\xi^{in}_{l,\gamma\delta}({\bf r'},t')\rangle
&=& 2T_g\left[\zeta+\frac{2\eta(d-1)}{d}\right]\delta_{\alpha\beta}\delta_{\gamma\delta}
\delta(t-t')\delta({\bf r}-{\bf r'}), \nonumber \\
\label{a.7}
\end{eqnarray}
and

\begin{equation}
\langle\xi^{ex}_\perp({\bf r},t)\xi^{ex}_\perp({\bf r'},t')\rangle=
\langle\xi^{ex}_l({\bf r},t)\xi^{ex}_l({\bf r'},t')\rangle
=\frac{1}{n}\frac{2 \gamma_bT_b}{m}\delta(t-t')\delta({\bf r}-{\bf r'}).
\label{a.8}
\end{equation}
Finally, taking the Fourier transform of Eqs.~(\ref{a.6},~\ref{a.7},~\ref{a.8}) we obtain
Eq.~(\ref{fh.2}) and followings.

\section*{References}

\bibliographystyle{unsrt}
\bibliography{fluct.bib}

\end{document}